\documentclass{ws-procs9x6}
\usepackage{times}
\usepackage{psfrag}
\usepackage{color}
\usepackage{amssymb,graphicx}
\begin{document}

\newcommand{\fig}[2]{\includegraphics[width=#1]{./figures/#2}}
\newcommand{\Fig}[1]{\includegraphics[width=\columnwidth]{./figures/#1}}
\newcommand{\pfig}[2]{\parbox{#1}{\includegraphics[width=#1]{./figures/#2}}}

\newcommand{\sgn}{{\mathrm{sgn}}}
\newcommand{\rme}{{\mathrm{e}}}
\newcommand{\rmd}{{\mathrm{d}}}
\newcommand{\E}{\epsilon}
\newcommand{\R}{\mathbb{R}}
\newcommand{\N}{\mathbb{N}}
\newcommand{\half}{\frac12}
\newcommand{\ts}{\hskip0.1ex\raisebox{-1ex}[0ex][0.8ex]{\rule{0.1ex}{2.75ex}\hskip0.2ex}}

\title{How to measure the effective action for disordered systems}

\author{Kay J\"org Wiese and Pierre Le Doussal}

\address{Laboratoire de Physique Th\'eorique de l'Ecole Normale
Superieure, 24 rue Lhomond, 75005 Paris, France.}

\begin{abstract}
In contrast to standard critical phenomena, disordered systems need to be treated via the Functional Renormalization Group. The latter leads to a coarse grained disorder landscape, which after a finite renormalization becomes non-analytic, thus overcoming the predictions of the seemingly exact dimensional reduction. We review recent progress on how the non-analytic effective action can be measured both in simulations and experiments, and confront theory with numerical work. 
\end{abstract}


\bodymatter

\section{Introduction}\label{intro}

When talking to his experimental colleagues about the marvels of field theory and his recent achievements in computing the effective action of his favorite model, the conversation 
is likely to resemble this:

\noindent\underline{Theorist}: I have a wonderful field theory,  I can even calculate the effective action! 

\noindent\underline{Experimentalist}:
Can I see it in an experiment? Can I measure it? 

\noindent\underline{Theorist}: \ldots well, that's difficult, but it tells you all you want to know \ldots

\noindent\underline{Experimentalist}: okay, I understand, another of these unverifiable predictions, \ldots 

Here we will see how to measure it, considering the explicit and far-from-trivial example of elastic manifolds in a disordered environment. 
Due to a lack of space, we will not be able to give all arguments in the necessary details. We recommend that the reader consults the recent ``Basic Recipes and Gourmet Dishes''\cite{WieseLeDoussal2006}, to which we also refer for a more complete list of references.

\section{The disordered systems treated here -- our model}

\begin{figure}[b]
\centerline{\fig{0.25\textwidth}{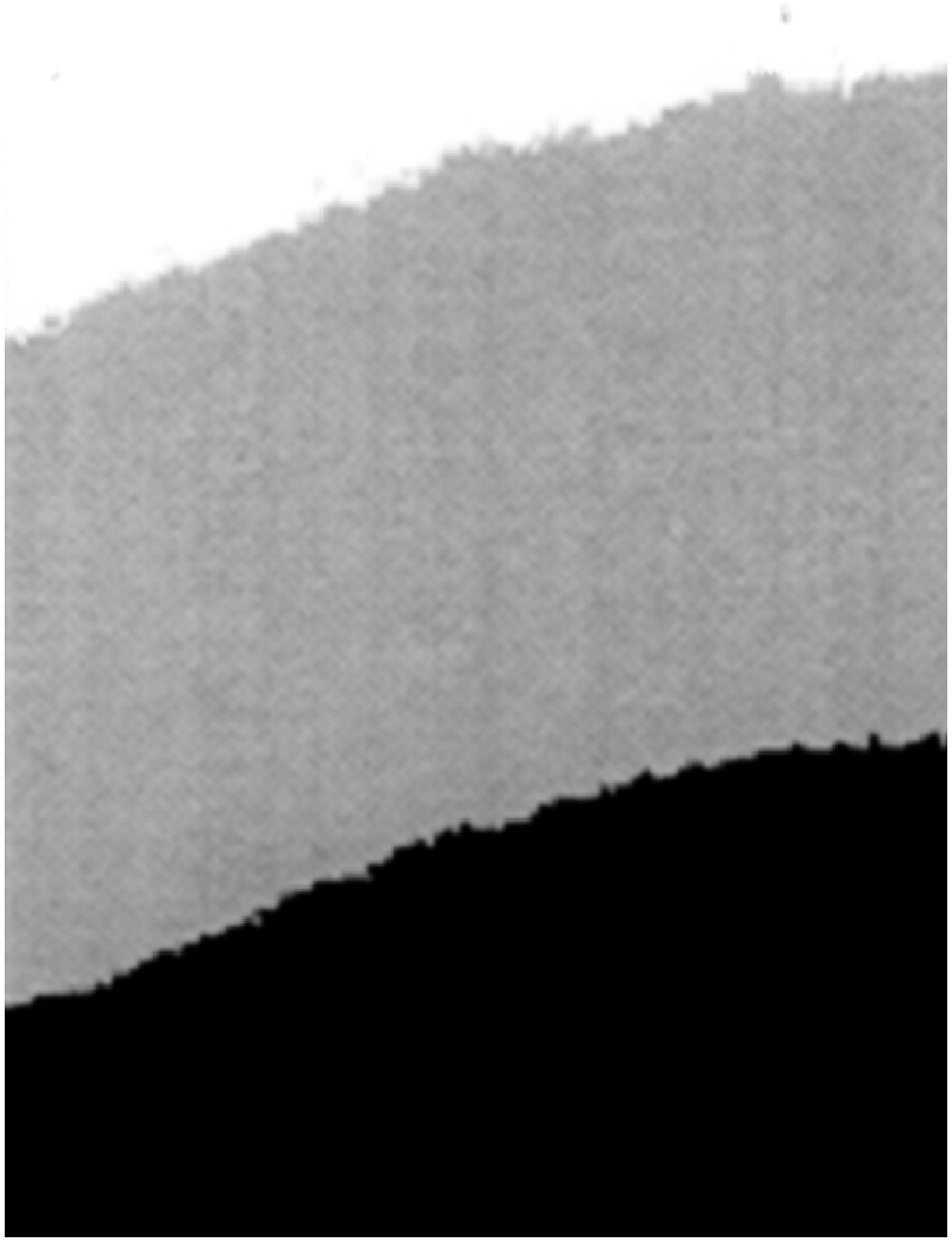}~~~\fig{0.7\textwidth}{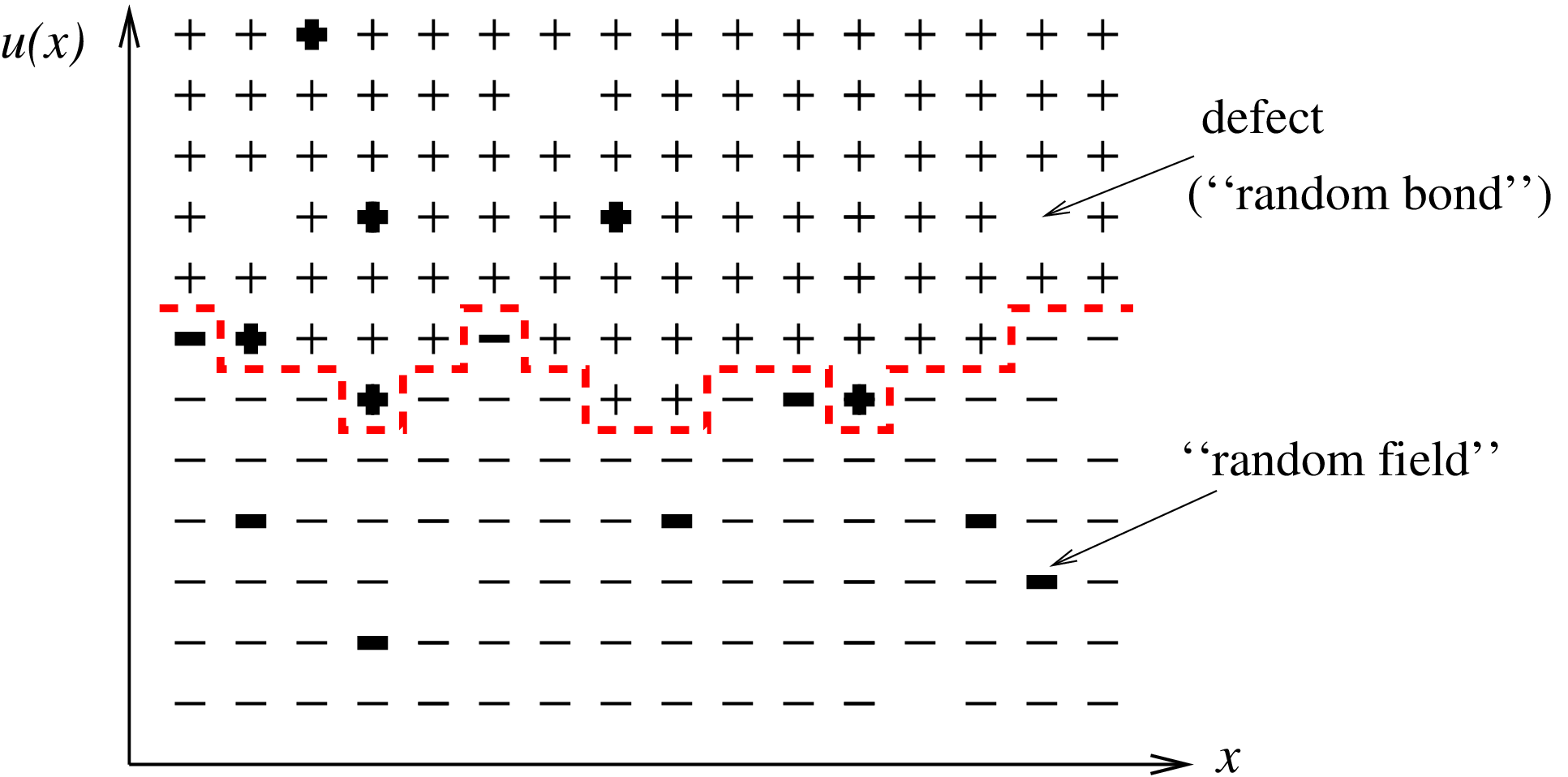}}
\caption{An Ising magnet at low temperatures forms a domain wall
described by a function $u (x)$ (right). An experiment on a thin
Cobalt film (left)
\protect\cite{LemerleFerreChappertMathetGiamarchiLeDoussal1998}; with
kind permission of the authors.}
\label{exp:Magnet}
\end{figure}
Let us first give some physical realizations. The simplest one
is an Ising magnet. Imposing boundary conditions with all spins up at
the upper and all spins down at the lower boundary (see figure 1), at
low temperature $T$, a domain wall will form in between. In a pure system at $T=0$, this
domain wall is completely flat; it will be  roughened by disorder. 
Two types of disorder are common:
random bond (which on a course-grained level represents missing
spins) and random field (coupling of the spins to an external random
magnetic field). Figure 1 shows, how the domain wall is described by a
displacement field $u (x)$.  Another example is the contact line of
water (or liquid hydrogen), wetting a rough substrate.  A realization
with a 2-parameter field $\vec{u} (\vec x) $ is the
deformation of a vortex lattice: the position of each vortex is deformed from $\vec x$ to
$\vec x+ \vec u (\vec x)$.  A 3-dimensional
example are charge density waves.

All these models are described
by a displacement field
\begin{equation}\label{u}
x\in \R^d \ \longrightarrow\  \vec u (x) \in \R^N
\ .
\end{equation}
For simplicity, we now set $N=1$.  After some initial
coarse-graining, the energy ${\cal H}={\cal H}_{\mathrm{el}}+{\cal
H}_{\mathrm{DO}}$ consists out of two parts: the elastic energy, and the disorder:
\begin{equation}
{\cal H}_{\mathrm{el}}[u] = \int \rmd ^d x \, \half \left( \nabla u
(x)\right)^2\ , \qquad {\cal H}_{\mathrm{DO}}[u] = \int \rmd ^{d} x \, V (x,u (x))\ .
\end{equation}
In order to proceed, we need to specify the  correlations of
disorder\cite{review}:
\begin{equation}\label{DOcorrelR}
\overline{V (x,u)V (x',u')} := \delta ^{d } (x-x') R (u-u')
\ .
\end{equation}
Fluctuations $u$ in
the transversal direction will scale as
\begin{equation}\label{roughness}
\overline{\left[u (x)-u (y) \right]^{2}}  \sim  |x-y|^{2\zeta }\ .
\end{equation}
There are several useful observables. We already introduced the
roughness-exponent $\zeta $. The second is the renormalized
(effective) disorder function $R(u)$, and it is this object we want to measure here. 
Introducing replicas and averaging over disorder, 
we can write down the bare action or  {\em replica-Hamiltonian}
\begin{equation}\label{H}
{\cal H}[u] = \frac{1}{T} \sum _{a=1}^{n}\int \rmd ^{d }x\, \half
\left(\nabla u_{a} (x) \right)^{2} -\frac{1}{2 T^{2}}  \sum
_{a,b=1}^{n} \int \rmd ^{d }x\, R (u_{a} (x)-u_{b} (x))\ .
\end{equation}
Let us stress that one could alternatively pursue a dynamic  or a
supersymmetric formulation. Since our treatment is perturbative in $R(u)$, the result is unchanged.

\section{Dimensional reduction}\label{dimred} There is a beautiful and rather
mind-boggling theorem relating disordered systems to pure systems
(i.e.\ without disorder), which applies to a large class of systems,
e.g.\ random-field systems and elastic manifolds in disorder. It is
called dimensional reduction and reads as
follows\cite{EfetovLarkin1977}:

\noindent {\underline{Theorem:}} {\em A $d$-dimensional disordered
system at zero temperature is equivalent to all orders in perturbation
theory to a pure system in $d-2$ dimensions at finite temperature. }

Experimentally, one finds that this result is wrong, the question being why? 
Let us stress that there are no missing diagrams or any such
thing, but that the problem is more fundamental: As we will see later,
the proof makes assumptions, which are not satisfied.  
Before we try to understand why this is so and how
to overcome it, let us give one more example.  We know that the width $u$ of a $d$-dimensional manifold at
finite temperature in the absence of disorder scales as $u\sim
x^{(2-d)/2}$. Making the dimensional shift implied by dimensional
reduction leads to
\begin{equation}\label{zetaDR}
\overline{\left[ u (x)-u (0) \right]^{2}} \sim x^{4-d} \equiv x^{2\zeta }
\quad \mbox{i.e.}\quad \zeta =\frac{4-d}{2}\ .
\end{equation}

\section{The Larkin-length}\label{Larkin}
To understand the failure of dimensional reduction, let us turn to an
interesting argument given by Larkin \cite{Larkin1970}. He considers a
piece of an elastic manifold of size $L$. If the disorder has
correlation length $r$, and characteristic potential energy $\bar f$,
this piece will typically see a potential energy of strength
$
E_{\mathrm{DO}} = \bar f \left(\frac{L}{r} \right)^{\!\frac{d}{2}}.$
On the other hand, there is an elastic energy, which scales like
$E_{\mathrm{el}} = c\, L^{d-2}.$
These energies are balanced at the  {\em Larkin-length} $L=L_{c}$
with $L_{c} = \left(\frac{c^{2}}{\bar f^{2}}r^{d} \right)^{\frac{1}{4-d}}.$
More important than this value is the observation that in all
physically interesting dimensions $d<4$, and at scales $L>L_{c}$, the
membrane is pinned by disorder; whereas on small scales the elastic energy
dominates. Since the disorder has a lot of minima which are far apart
in configurational space but close in energy (metastability), the
manifold can be in either of these minimas, and the ground-state is no
longer unique. However exactly this is assumed in the proof of
dimensional reduction.

\section{The functional renormalization group (FRG)}\label{FRG}

Let us now discuss a way out of the dilemma: Larkin's argument
suggests that $d=4$ is the upper critical
dimension. So we would like to make an $\epsilon =4-d$ expansion. On
the other hand, dimensional reduction tells us that the roughness is
$\zeta =\frac{4-d}{2}$ (see (\ref{zetaDR})). Even though this is
systematically wrong below four dimensions, it tells us correctly that
at the critical dimension $d=4$, where disorder is marginally
relevant, the field $u$ is dimensionless. This means that having
identified any relevant or marginal perturbation (as the disorder), we
can find another such perturbation by adding more powers of
the field. We can thus not restrict ourselves to keeping solely the
first moments of the disorder, but have to keep the whole
disorder-distribution function $R (u)$. Thus we need a {\em functional
renormalization group} treatment (FRG). Functional renormalization is
an old idea, and can e.g.\ be found in
\cite{WegnerHoughton1973}.  For disordered systems, it was first
proposed in 1986 by D.\ Fisher \cite{DSFisher1986}.  Performing an
infinitesimal renormalization, i.e.\ integrating over a momentum shell
\`a la Wilson, leads to the flow $\partial _{\ell} R (u)$, with
($\epsilon =4-d$)
\begin{equation}\label{1loopRG}
\partial _{\ell} R (u) = \left(\epsilon -4 \zeta  \right) R (u) +
\zeta u R' (u) + \frac{1}{2} R'' (u)^{2}-R'' (u)R'' (0)\ .
\end{equation}
The first two terms come from the rescaling of $R$  and $u$  respectively. The last two terms are the
result of the 1-loop calculations, see e.g.\ \cite{DSFisher1986, WieseLeDoussal2006}.

\begin{figure}[t]
\centerline{\Fig{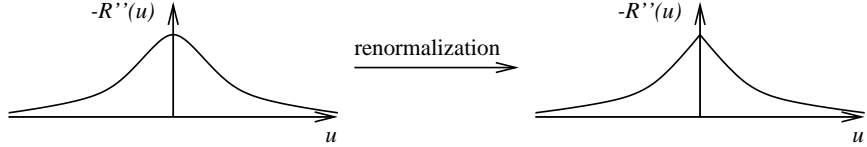}}
\caption{Change of $-R'' (u)$ under renormalization and formation of
the cusp.} \label{fig:cusp}
\end{figure}

More important than the form of this equation is it actual solution,
sketched in figure \ref{fig:cusp}.
After some finite renormalization, the second derivative of the
disorder $R'' (u)$ acquires a cusp at $u=0$; the length at which this
happens is the Larkin-length. How does this overcome dimensional
reduction?  To understand this, it is interesting to study the flow of
the second and forth moment. Taking derivatives of (\ref{1loopRG})
w.r.t.\ $u$ and setting $u$ to 0, we obtain
\begin{eqnarray}
\partial_{\ell} R'' (0) &=& \left(\epsilon -2 \zeta  \right) R'' (0) +
R''' (0)^{2} \ \longrightarrow \ \left(\epsilon -2 \zeta  \right) R''
(0)\label{R2of0}\\\nonumber
\partial_{\ell} R'''' (0) &=& \epsilon  R'''' (0) + 3 R'''' (0)^{2} +4 R'''
(0)R''''' (0)  \longrightarrow\epsilon  R'''' (0) + 3 R''''
(0)^{2}\label{R4of0}.
\end{eqnarray}
Since $R (u)$ is an even function, and moreover the microscopic
disorder is smooth (after some initial averaging, if necessary), $R'''
(0)$ and $R''''' (0)$ are 0, which we have already indicated. The above equations for $R'' (0)$
and $R'''' (0)$ are in fact closed. The first tells us first
that the flow of $R'' (0)$ is trivial and that $\zeta =\epsilon
/2\equiv \frac{4-d}{2}$. This is exactly the result predicted by
dimensional reduction. The appearance of the cusp can be inferred from the second one. Its solution is
$
R'''' (0)\ts _{\ell}= \frac{c\,\rme^ {\epsilon \ell }}{1-3\, c \left(\rme^
{\epsilon \ell} -1 \right)/ \epsilon }$, with $c= R'''' (0)\ts _{\ell=0}
$.
Thus after a finite renormalization $R'''' (0)$ becomes infinite: The
cusp appears. By analyzing the solution of the flow-equation
(\ref{1loopRG}), one also finds that beyond the Larkin-length $R''
(0)$ is no longer given by (\ref{R2of0}) with $R''' (0)^{2}=0$.  The
correct interpretation of (\ref{R2of0}), which remains valid after the
cusp-formation, is 
$
\partial_{\ell} R'' (0) = \left(\epsilon -2 \zeta  \right) R'' (0)  +R''' (0^{+})^{2} \label{R2of0after}.
$
Renormalization of the whole function thus overcomes dimensional
reduction.  The appearance of the cusp also explains why dimensional
reduction breaks down: The simplest way to see this is by redoing the
proof for elastic manifolds in disorder, which in the absence of
disorder is a simple Gaussian theory. Terms contributing to the
2-point function involve $R'' (0)$, $TR'''' (0)$ and higher
derivatives of $R (u)$ at $u=0$, which all come with higher powers of
$T$. To obtain the limit of $T\to 0$, one sets $T=0$, and only $R''
(0)$ remains. This is the dimensional-reduction result. However we
just saw that $R'''' (0)$ becomes infinite. Not surprisingly
$R'''' (0) T$ may also contributes; indeed one  can show that it does,
hence the proof fails.

\enlargethispage{.1mm}

\section{The cusp and shocks}\label{shocks}\begin{figure}[t]
\centerline{\raisebox{10mm}[0mm][0mm]{\parbox{0in}{(a)}}\pfig{.4\textwidth}{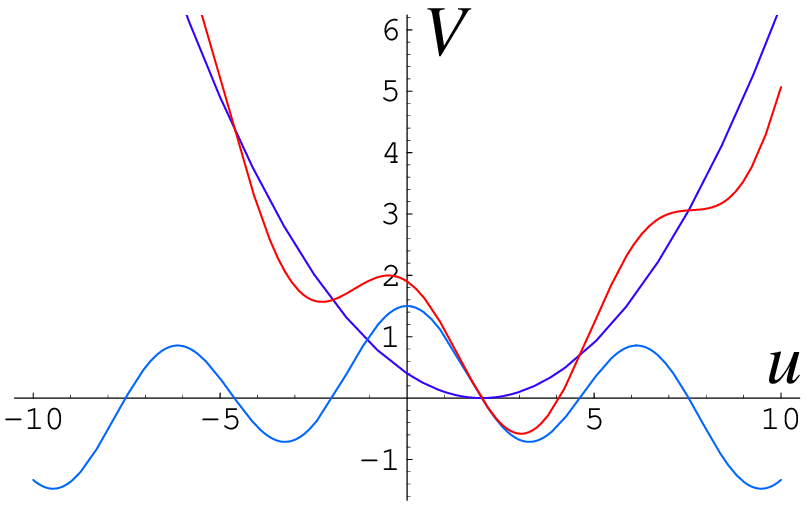}
$\quad \stackrel{\mbox{\small
minimize}}{-\!\!\!-\!\!\!-\!\!\!-\!\!\!\longrightarrow}
\quad$
\raisebox{0mm}[0mm][0mm]{\parbox{0in}{(b)}}\pfig{.4\textwidth}{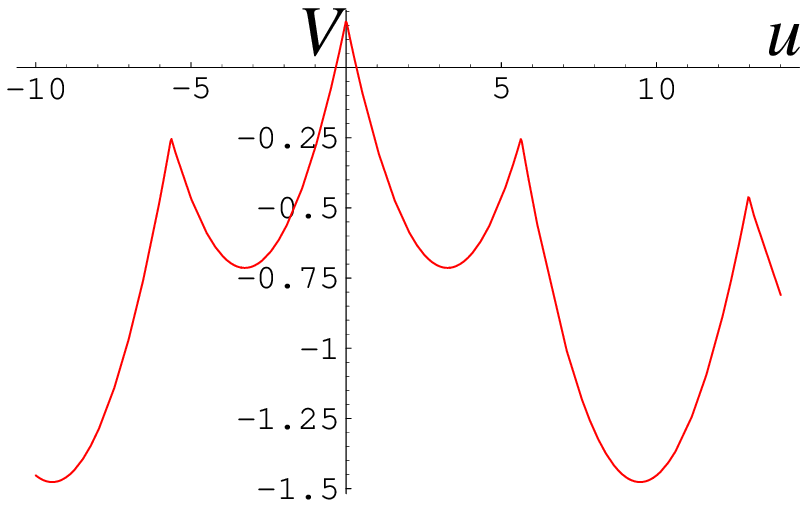}}
\centerline{\raisebox{10mm}[0mm][0mm]{\parbox{0in}{(c)}}\pfig{.4\textwidth}{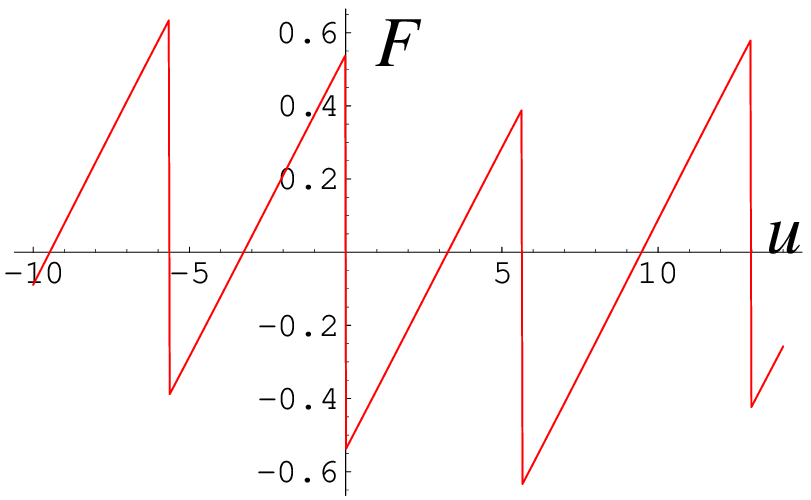}
$\quad \stackrel{\mbox{\small average}}
{-\!\!\!-\!\!\!-\!\!\!-\!\!\!\longrightarrow} \quad$
\raisebox{0mm}[0mm][0mm]{\parbox{0in}{(d)}}\pfig{.4\textwidth}{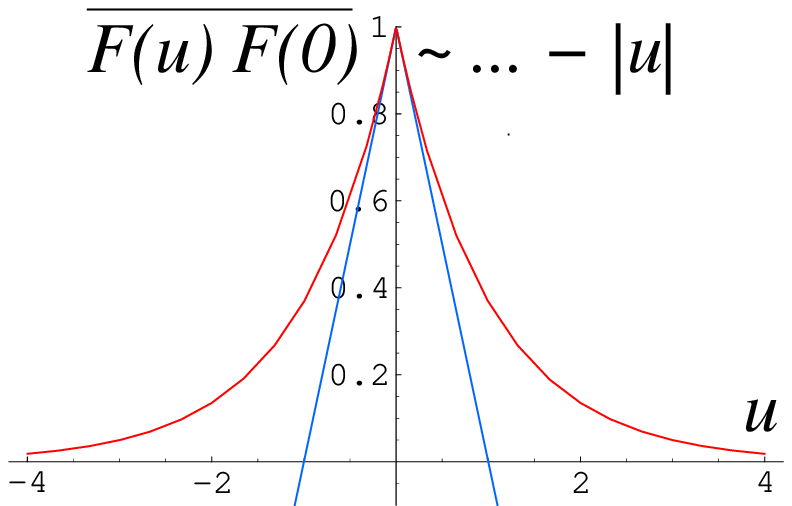}}
\caption{Generation of the cusp, as explained in the main text.} \label{minfig}
\end{figure}Let us give a simple
argument of why a cusp is a physical necessity, and not an
artifact. The argument is quite old and appeared probably first in the
treatment of correlation-functions by shocks in Burgers
turbulence. It was nicely illustrated in \cite{BalentsBouchaudMezard1996}. 
Suppose, we want to integrate out a single degree of freedom coupled with a spring. 
This harmonic potential and the disorder term are
represented by the parabola and the lowest curve on figure
\ref{minfig}(a) respectively; their sum is the remaining curve.  For a
given disorder realization, the minimum of the potential as a function
of $u$ is reported on figure \ref{minfig}(b). Note that it has
non-analytic points, which mark the transition from one minimum to
another.  Taking the derivative of the potential leads to the force in
figure \ref{minfig}(c). It is characterized by almost linear pieces,
and shocks (i.e.\ jumps).  Calculating the force-force correlator, the
dominant contribution for small distances is due to
 shocks. Their contribution is proportional to their
probability, i.e.\ to the distance between
the two observable points. This leads to $\overline{F (u)F (0)} = \overline{F
(0)^{2}} - c |u|$, with some numerical coefficient $c$.

\section{The field-theoretic version}\label{measurecusp}

The above toy model can be generalized to the field theory \cite{LeDoussal2006b}. 
Consider an  interface in a random potential, and add a
quadratic potential well, centered around $w$:
\begin{eqnarray}
{\cal H}_{\mathrm{tot}}^{w}[u] =  \int_{x}\frac{m^2}{2} (u(x)-w)^2 + {\cal
H}_{\mathrm{el}}[u] + {\cal H}_{\mathrm{DO}}[u]\ .
\end{eqnarray}
In each sample (i.e.\ disorder configuration), and given $w$, one
finds the minimum energy configuration. This 
ground state energy is
\begin{eqnarray}
\hat V(w) := \min_{u(x)} {\cal H}_{\mathrm{tot}}^{w}[u]\ .
\end{eqnarray}
It varies with $w$ as well as from sample to sample. Its second
cumulant
\begin{eqnarray}
\overline{ \hat V(w)  \hat V(w') }^c = L^d R(w-w') \label{defR}
\end{eqnarray}
defines a function $R(w)$ which is proven \cite{LeDoussal2006b} to
be the same function computed in the field theory, defined from the zero-momentum
effective action \cite{LeDoussalWieseChauve2003}.
Physically, the role of the well is to forbid the interface to wander
off to infinity. The limit of small $m$ is taken to reach the
universal limit. The factor of volume $L^d$ is necessary, since the width
$\overline{u^2}$ of the interface in the well cannot grow much more
than $m^{-\zeta}$. This means that the interface is made of roughly
$L/L_m$ pieces of internal size $L_m \approx m$ pinned independently:
(\ref{defR}) expresses the central-limit theorem and $R(w)$
measures the second cumulant of the disorder seen by each piece.

\begin{figure}[t]
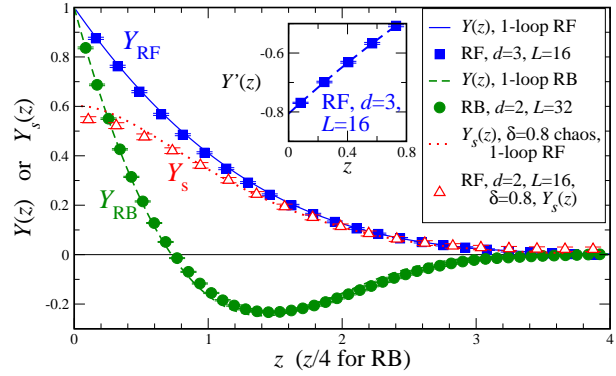
\setlength{\unitlength}{1.4mm}
\fboxsep0mm \centerline{\mbox{\fig{8cm}{compareRFRBchaos}}}
\caption{Filled symbols show numerical results for $Y(z)$, a
normalized form of the interface displacement correlator $-R''(u)$
[Eq.\ (\ref{defDe})], for $D=2+1$ random field (RF) and $D=3+1$ random
bond (RB) disorders. These suggest a linear cusp. The inset plots the
numerical derivative $Y'(z)$, with intercept $Y'(0)\approx -0.807$
from a quadratic fit (dashed line).  Open symbols plot the
cross-correlator ratio $Y_s(z)=\Delta_{12}(z)/\Delta_{11}(0)$ between
two related copies of RF disorder. It does not exhibit a cusp.  The
points are for confining wells with width given by $M^2=0.02$.
Comparisons to 1-loop FRG predictions (curves) are made with no
adjustable parameters. Reprinted from \cite{MiddletonLeDoussalWiese2006}.}  
\label{f:Alan1}
\end{figure}
The nice thing about (\ref{defR}) is that it can be measured. One
varies $w$ and computes (numerically) the new ground-state energy;
finallying averaging over many realizations. This has been performed
recently in \cite{MiddletonLeDoussalWiese2006} using a powerful
exact-minimization algorithm, which finds the ground state in a time
polynomial in the system size. In fact, what was measured there are
the fluctuations of the center of mass of the interface $u(w)=L^{-d}
\int \rmd^d x\, u_0(x;w)$:
\begin{eqnarray}
\overline{[w-u(w)] [w'-u(w')] }^c = m^{-4} L^{-d} \Delta(w-w')
\label{defDe}
\end{eqnarray}
which measures directly the correlator of the pinning force
$\Delta(u)=-R''(u)$. To see why it is the total force, write the
equilibrium condition for the center of mass $m^2 [w-u(w)] + L^{-d}
\int \rmd^d x\, F(x,u)=0$ (the elastic term vanishes if we use periodic
b.c.). The result is represented in figure \ref{f:Alan1}. It is most
convenient to plot the function $Y=\Delta(u)/\Delta(0)$ and normalize
the $u$-axis to eliminate all non-universal scales.
\begin{figure}[b]
\mbox{\fig{0.50\textwidth}{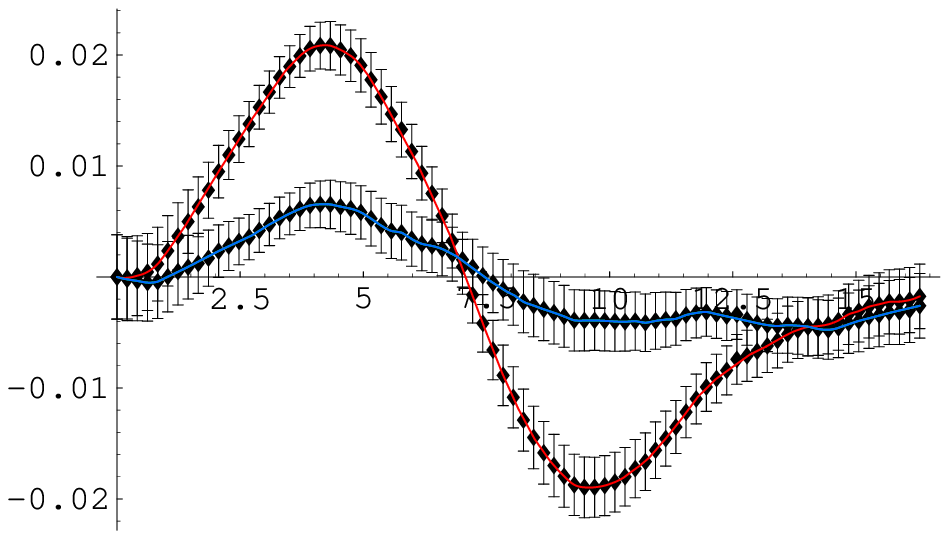}}~~\mbox{\fig{0.5\textwidth}{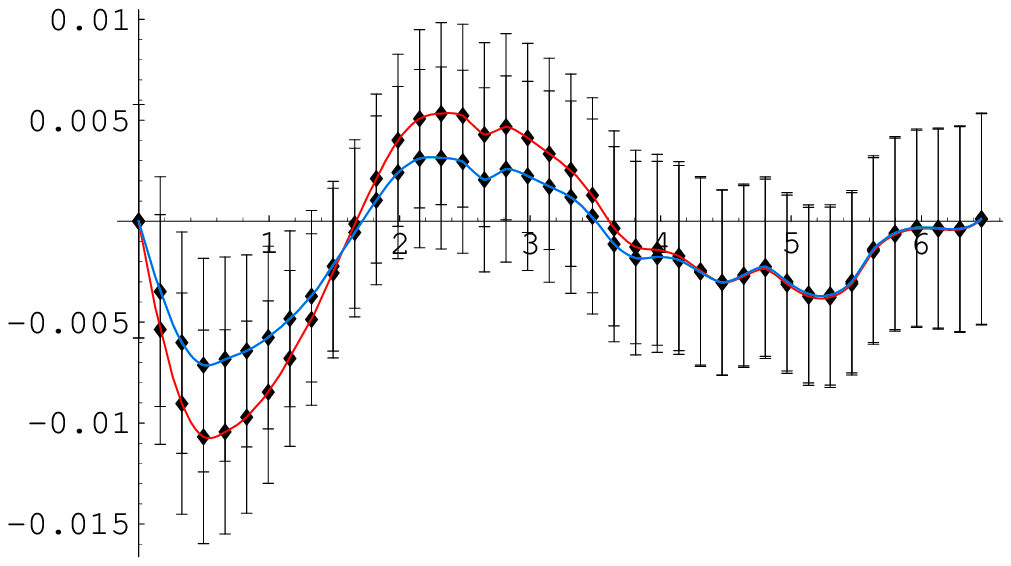}}
\caption{The measured $Y(u)$ with the 1- and 2-loop corrections subtracted. Left: RB-disorder, right: RF-disorder. One sees that the 2-loop corrections improve the precision.}
\end{figure}
\begin{figure}
\setlength{\unitlength}{.37mm}
\mbox{\begin{picture}(300,155)
\put(0,0){\fig{100mm}{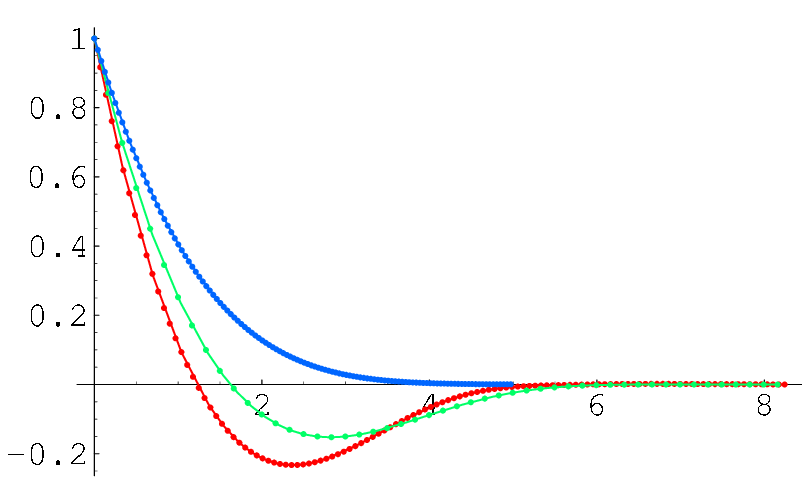}}
\put(0,150){ $ \Delta (u) $}
\put(260,20){ $u$} 
\put(35,18){\color{red} $m^{2}=1$}
\put(144,18){\color{green} $m^{2}=0.5$}
\put(92,52){\color{blue} $m^{2}=0.003$}
\put(150,45){\fig{0.5\textwidth}{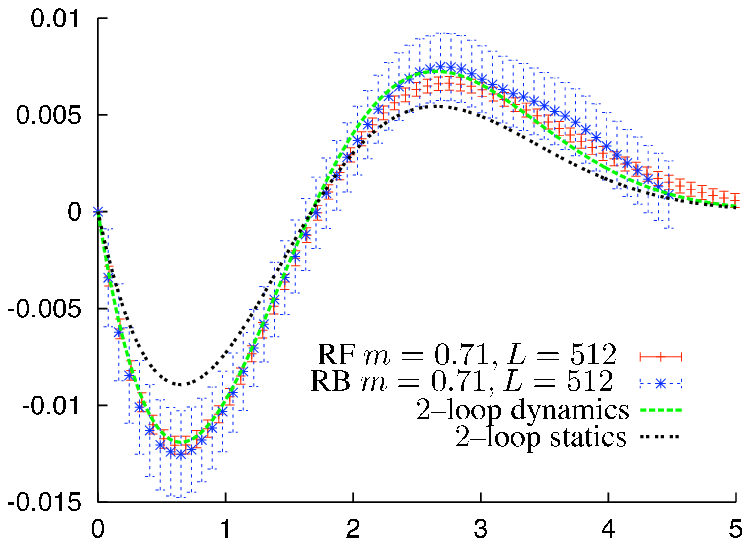}}
\end{picture}}
\caption{{Running the RG in a numerical simulation: Crossover from RB disorder to RF for a driven particle (left)\cite{MiddletonLeDoussalWiese2006}. Residual error for $Y(u)$ for a driven string \cite{RossoLeDoussalWiese2006a} which show that statics and depinning are controled by different
fixed points.}}
\label{f:dyn}
\end{figure}
The plot in figure \ref{f:Alan1} is free of any parameter. It has
several remarkable features. First, it clearly shows that a linear
cusp exists in any dimension. Next it is very close to the 1-loop
prediction. Even more remarkably the statistics is good enough \cite{MiddletonLeDoussalWiese2006} to reliably compare the deviations to the 2-loop predictions of \cite{ChauveLeDoussalWiese2000a}.

When we vary the position $w$ of the
center of the well, it is not a real motion. It  means to find the
new ground state for each $w$.  Literally ``moving'' $w$ is another
interesting possibility: It measures the universal properties of the so-called ``depinning transition''
\cite{LeDoussalWiese2006a,RossoLeDoussalWiese2006a}. This was recently
implemented numerically (see Fig.~\ref{f:dyn}).

\subsection*{Acknowledgments}\label{ack} It is a pleasure to thank
Wolfhard Janke and Axel Pelster, the organizers of PI-2007 for the opportunity to give this
lecture. We thank Alan Middleton and Alberto Rosso for fruitful collaborations which allowed
to measure the FRG effective action in numerics and Andrei Fedorenko and Werner Krauth for stimulating discussions. This work has been supported by
ANR (05-BLAN-0099-01).

\bibliography{../../citation/citation}

\bibliographystyle{ws-procs9x6}

\end{document}